\documentclass[%
 reprint,
superscriptaddress,
 amsmath,amssymb,
 aps,
prl,
]{revtex4-1}
\usepackage[utf8]{inputenc}
\usepackage[T1]{fontenc}
\usepackage{placeins}
\usepackage[normalem]{ulem}
\usepackage{soul}
\usepackage{bibunits}
\usepackage{tipa}
\usepackage{dsfont}
\usepackage{physics}
\usepackage{graphicx}
\makeatletter

\begin{document}
\defaultbibliographystyle{apsrev4-1}
\begin{bibunit}

\title{Ab initio calculations of neutrinoless $\beta \beta$ decay refine neutrino mass limits}

\author{A.~Belley}
\affiliation{TRIUMF, Vancouver, BC, Canada}
\affiliation{Department of Physics \& Astronomy, University of British Columbia, Vancouver, BC, Canada}
\author{T.~Miyagi}
\affiliation{Technische Universit\"at Darmstadt, Department of Physics, 64289 Darmstadt, Germany}
\affiliation{ExtreMe Matter Institute EMMI, GSI Helmholtzzentrum f\"ur Schwerionenforschung GmbH, 64291 Darmstadt, Germany}
\affiliation{Max-Planck-Institut f\"ur Kernphysik, Saupfercheckweg 1, 69117 Heidelberg, Germany}
\author{S.~R.~Stroberg}
\affiliation{Department of Physics and Astronomy, University of Notre Dame, Notre Dame, IN 46556 USA}
\author{J.~D.~Holt}
\affiliation{TRIUMF, Vancouver, BC, Canada}
\affiliation{Department of Physics, McGill University, Montr\'eal, QC, Canada}

\date{\today}

\begin{abstract}

Neutrinos are perhaps the most elusive known particles in the universe. 
We know they have some nonzero mass, but unlike all other particles, the absolute scale remains unknown. 
In addition, their fundamental nature is uncertain; they can either be their own antiparticles or exist as distinct neutrinos and antineutrinos.
The observation of the hypothetical process of neutrinoless double-beta ($0\nu\beta\beta$) decay would at once resolve both questions, while providing a strong lead in understanding the abundance of matter over antimatter in our universe~\cite{Fukugita1986}. 
In the scenario of light-neutrino exchange, the decay rate is governed by, and thereby linked to the effective mass of the neutrino via, the theoretical nuclear matrix element (NME).
In order to extract the neutrino mass, if a discovery is made, or to assess the discovery potential of next-generation searches, it is essential to obtain accurate NMEs for all isotopes of experimental interest. 
However, two of the most important cases, $^{130}$Te and $^{136}$Xe, lie in the heavy region and have only been accessible to phenomenological nuclear models.
In this work we utilize powerful advances in ab initio nuclear theory to compute NMEs from the underlying nuclear and weak forces driving this decay, including the recently discovered short-range component~\cite{Cirigliano2021}. 
We find that ab initio NMEs are generally smaller than those from nuclear models, challenging the expected reach of future ton-scale searches as well as claims to probe the inverted hierarchy of neutrino masses~\cite{KamLand2023}.
With this step, ab initio calculations with theoretical uncertainties are now feasible for all isotopes relevant for next-generation $0\nu\beta\beta$ decay experiments.

\end{abstract}

\flushbottom
\maketitle

\thispagestyle{empty}

Lepton number is a conserved quantity in the standard model of particle physics; a process which creates more leptons than anti-leptons (i.e., creates matter with no antimatter) has never been observed and would have a tremendous impact on our understanding of the fundamental particles and forces that constitute our universe. 
Neutrinoless double-beta ($0\nu\beta\beta$) decay is the best known probe of lepton-number violation~\cite{Avig08RMP}, as it hypothetically changes two neutrons into two protons via emission of two electrons, but with no electron antineutrinos.  
The primary mode for lepton-number violation in this process is the annihilation, or exchange, of two light electron neutrinos produced in $\beta \beta$ decay, if and only if they are Majorana (i.e., their own antiparticles). 
Extensions of the standard model would potentially allow for the exchange of new heavy neutrinos, giving the possibility for $0\nu\beta\beta$ decay experiments to complement accelerators such as the Large Hadron Collider in searches for new particles in nature~\cite{Helo2013,Cirigliano2022}. 
To differentiate between models for such exotic  processes, observation in two or more isotopes, as well as reliable values of the NMEs, are needed. 

\begin{figure*}[t]
    \centering
    \includegraphics[width=\linewidth]{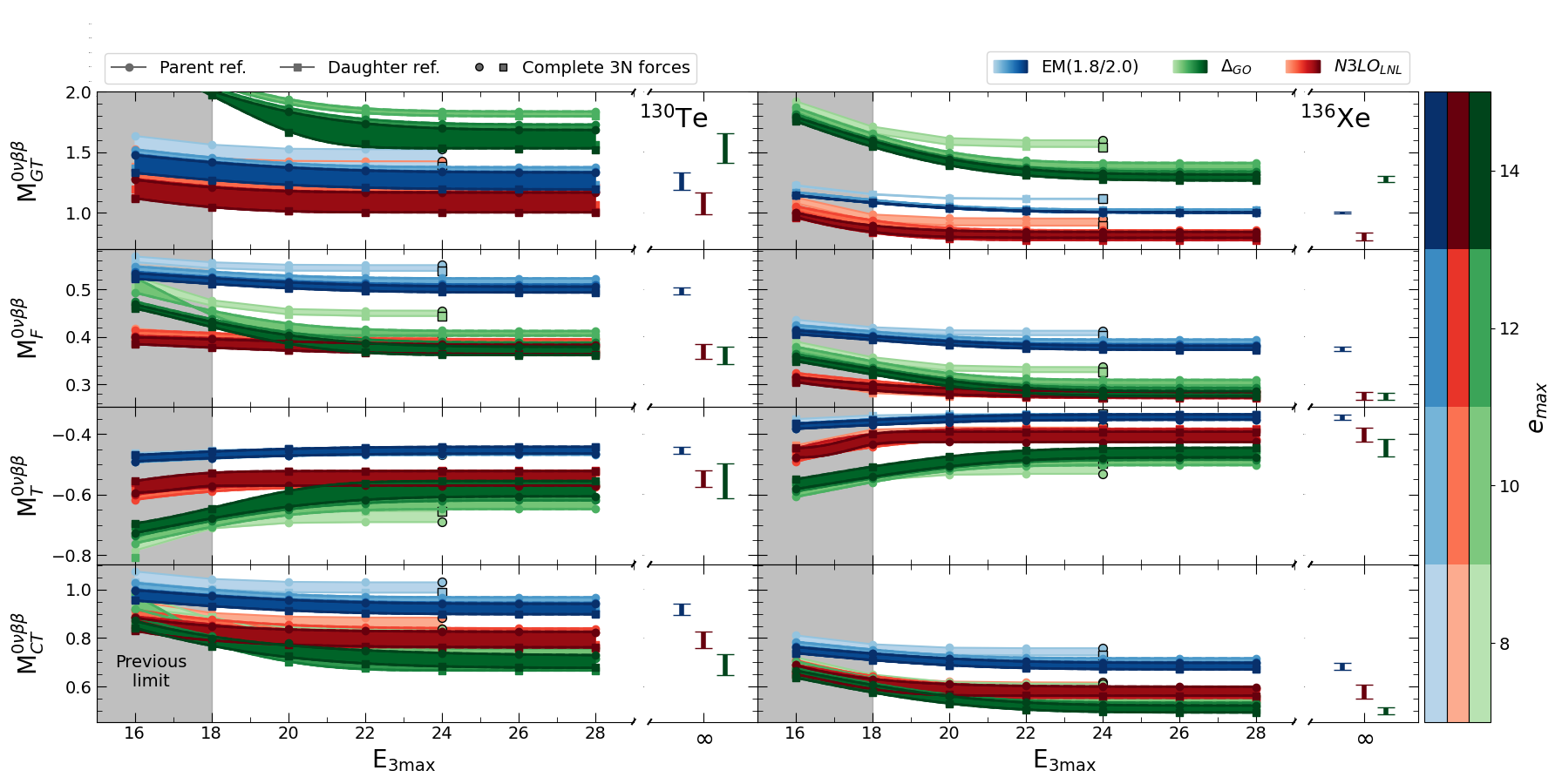}
    \caption{Convergence of the NMEs in both $e_{\textrm{max}}$ and $E_{3\textrm{max}}$ for each operator contributing to $0\nu\beta\beta$ decay, where the factors $-\big(\frac{g_V}{g_A}\big)^2$ and $- 2g_{\nu}^{NN}$ are included for the Fermi and contact terms, respectively. 
    Increases in the single-particle space, $e_{\textrm{max}}$, are shown with a color gradient for each interaction. 
    Values at infinity are extrapolated from points including all 3N forces using an exponential (see Methods for details). 
    The final error bar for extrapolated values represents both the spread from the choice of reference state as well as the error from both extrapolations. 
    }
    \label{fig:convergence}
\end{figure*}

At the moment, should an observation be made, the ability of experiments to identify the exact lepton-number-violating mechanism involved in $0\nu\beta\beta$ decay would be greatly hindered by our poor knowledge of the NMEs~\cite{Graf2022}. 
Furthermore, comparing the relative sensitivities to new physics of proposed next-generation experiments is also difficult. 
Since different candidate isotopes are typically used for different searches, the NME for each isotope is needed to convert expected experimental half-life limits to a universal measure of the sensitivity to the new process.
The NMEs are consequently vital inputs for guiding the funding, strategic planning, and ultimate design of future experiments worldwide.

An accurate calculation of the NMEs has been a challenge to nuclear theory for more than 40 years, as it primarily requires a consistent treatment of both nuclear and electroweak forces. 
In addition, all key experimental isotopes for worldwide searches are open-shell heavy systems, where the nuclear many-body problem has historically been difficult, if not impossible, to treat accurately. 
To date, all existing calculations have originated from phenomenological nuclear models, where the lack of constraining data has resulted in a fairly large spread between calculations~\cite{Engel2017}. 
These models all neglect some essential physics, and furthermore, do not have a clear way to robustly assess systematic uncertainties, significantly inhibiting our ability to interpret experimental limits in terms of excluded neutrino masses.

First-principles, or ab initio, nuclear theory is founded on nuclear interactions rooted in quantum chromodynamics (QCD), namely, those derived within chiral effective field theory (EFT)~\cite{Epelbaum2009, Machleidt2011}.
Chiral EFT is a systematic low-energy expansion for both the nuclear and electroweak interactions involved in $0\nu\beta\beta$ decay. 
At this low-energy scale, long-range forces are mediated by exchange of pions, while information from unresolved high-energy degrees of freedom is encoded in short-range contact interactions, whose couplings are typically constrained to data in few-nucleons systems. 
The nuclear many-body problem is then approximately solved within some systematically improvable, nonperturbative framework such as the no-core shell-model~\cite{Barrett2013, Navratil2016}, quantum Monte Carlo~\cite{Carlson2015,Lynn2019}, coupled-cluster theory~\cite{Hage10RPP}, or the in-medium similarity renormalization group (IMSRG) used in this study~\cite{Hergert2020}.

As a gateway to $0\nu\beta\beta$ decay, ab initio studies have successfully resolved the decades-old ``quenching" problem in single-beta decays \cite{Gysbers2019}, where nuclear models had failed to predict decay rates, due to missing many-body correlations and two-body weak currents.
Furthermore, several ab initio methods, including ours, have made first steps in NME calculations for both fictitious decays and the lightest candidate isotope $^{48}$Ca~\cite{Yao20Bench,Yao2020,Belley2021,Novario2021}, and encouraging agreement is found within theoretical uncertainties. 
With the valence-space (VS) formulation of the IMSRG~\cite{Stroberg2017,Stro19ARNPS} (see Methods), we have also provided first results for NMEs in $^{76}$Ge and $^{82}$Se~\cite{Belley2021}. 
Until recently, the heavy nuclei $^{130}$Te and $^{136}$Xe, vital targets in several of the most prominent experimental searches, were well beyond the reach of ab initio theory, due to severe limitations on including effects of three-nucleon (3N) forces in large single-particle spaces. 
However, a recent breakthrough in how these matrix elements are stored and implemented~\cite{Miyagi2022} has extended the range of converged ab initio calculations to the $^{136}$Xe region, and even as far as $^{208}$Pb, where the neutron skin thickness has been linked to nucleon-nucleon (NN) scattering data \cite{Hu2022}.

\begin{figure}[t]
    \centering
    \includegraphics[width=\linewidth]{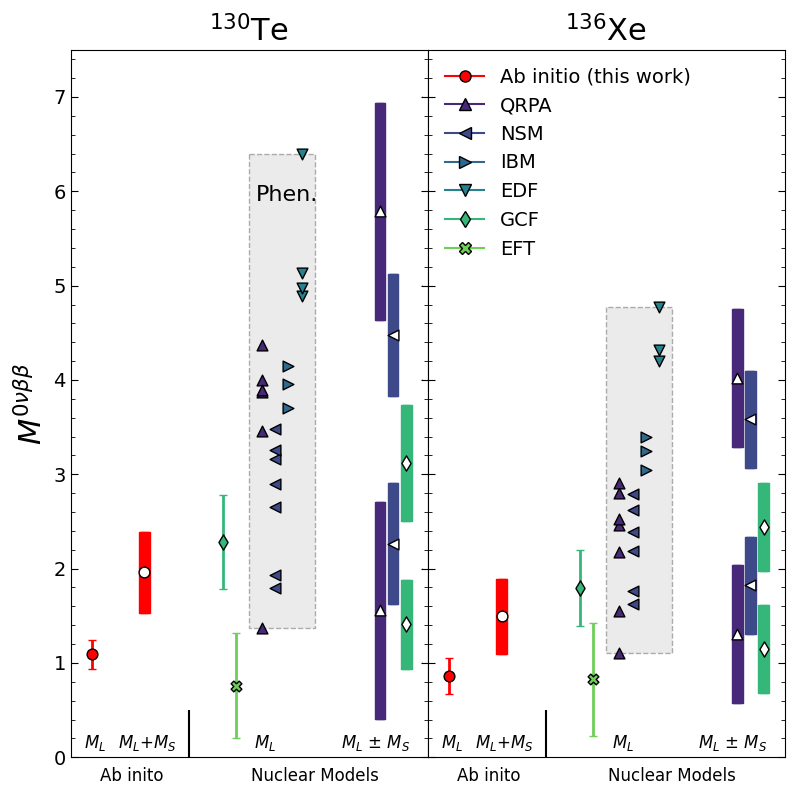}
    \caption{
    Range of ab initio VS-IMSRG results compared to nuclear models calculations of the NME, excluding (lines) and including (bands) the short-range contact term, denoted $M_L$ and $M_L$+$M_S$ respectively. 
    For nuclear models the sign of the short-range term is unknown, giving rise two possible bands.
    The box labelled ``Phen.'' represents the spread of phenomenological values typically used to interpret experimental results.
    }
    \label{fig:model_compare}
\end{figure}

An additional bottleneck for reliable NME calculations arises in the operator itself, where historically $0\nu\beta\beta$ decay has been expressed as the three long-range Gamow-Teller (GT), Fermi (F), and tensor (T) terms~\cite{Engel2017}. When considering this decay within an EFT framework, it was discovered that to properly renormalize the theory, a short-range contact operator (CT) must be promoted to leading order~\cite{Cirigliano2018a}:
\begin{equation*}
    M^{0\nu\beta\beta} = M^{0\nu\beta\beta}_{GT}-\big(\dfrac{g_V}{g_A}\big)^2M^{0\nu\beta\beta}_{F}+M^{0\nu\beta\beta}_{T} - 2g_{\nu}^{NN}M^{0\nu\beta\beta}_{CT}.
\end{equation*}
The associated coupling constant $g_{\nu}^{NN}$  was estimated to $30\%$ uncertainty by using a model to interpolate between low- and high-momentum contributions~\cite{Cirigliano2021} but should eventually be determined more precisely by matching to lattice QCD calculations.
Since this procedure requires matching in the two-nucleon sector to determine the coupling constant $g_{\nu}^{NN}$, it is only viable with ab initio approaches.
For the particular nuclear interactions used in this work, we take $g_{\nu}^{NN}$ from one such previous determination~\cite{Wirth2021} (a detailed description of each component is found in Methods).
We then make use of advances in nuclear structure calculations and include the contact term to provide ab initio results for the heaviest of the most prominent experimental isotopes, $^{130}$Te and $^{136}$Xe. 
We explore implications on existing and future searches as well as refined limits on the effective neutrino mass.

In Fig.~\ref{fig:convergence} we show the convergence of each operator contributing to the final NME, starting from three state-of-the-art parameterizations of chiral NN and 3N forces. 
Convergence must be reached for both the size of the single-particle space, denoted $e_{\textrm{max}}$, as well as the additional energy cut on included 3N forces, denoted $E_{3\textrm{max}}$ (see Methods for details).
As we show in the Extended Data and by the color gradients on Fig.~\ref{fig:convergence}, our results are converged to better than 2\% at $e_{\textrm{max}}=14$, i.e., 15 major harmonic-oscillator shells, so we focus the discussion here on $E_{3\textrm{max}}$. 
We illustrate this in Fig.~\ref{fig:convergence}, where all operators are well converged at $E_{3\textrm{max}} = 28$, while noting that for the previous limit of $E_{3\textrm{max}} = 18$, this is not the case for any NME component with any interaction.
In order to include all contributions from 3N forces, we would require $E_{3\textrm{max}} = 3 \cdot e_{\textrm{max}}$, but since this has not been achievable until recently for large $e_{\textrm{max}}$ values~\cite{Hebeler2023}, we instead use extrapolation techniques~\cite{Miyagi2022} to obtain values for full 3N forces at $e_{\textrm{max}}=10-14$. 
We then finally extrapolate the results including all 3N forces to an infinite model space size using an exponential fit.
Due to truncation of many-body operators in the IMSRG procedure (see Methods), our calculation depends on the choice of reference state (e.g., parent or daughter nucleus), which we also illustrate as bands in Fig.~\ref{fig:convergence}.

Taking the final results for all components together, we find the following NME values:
\begin{align*}
    ^{130}\textrm{Te}: M^{0\nu\beta\beta} \in [1.52, 2.40]\\
    ^{136}\textrm{Xe}: M^{0\nu\beta\beta} \in [1.08, 1.90].
\end{align*}
While the spread arises primarily from choice of nuclear interaction, we note it also includes reference-state dependence, basis extrapolation, the uncertainty coming from the closure approximation (see Methods), and the coefficient $g_{\nu}^{NN}$. 
While a rigorous statistical analysis is currently in progress using IMSRG-based emulators, we have recently observed that the NMEs are strongly correlated with the scattering phase shift in the $^{1}S_0$ (spin-singlet) partial wave. 
Since this quantity is very well reproduced by all interactions used in this work, we expect the spread given here to likely be representative of the final value of the NME.

In Fig.~\ref{fig:model_compare}, we compare our ab initio results to three other classes of calculations:~i) phenomenological nuclear models that do not include the short-range contributions; ii) phenomenological nuclear models that attempt to estimate the possible contributions of the short-range contact; and iii) an EFT approach that uses a possible correlation between $0\nu\beta\beta$ decay and the double Gamow-Teller charge exchange transition NMEs~\cite{Brase2022}. 
These phenomenological models have traditionally been used by experimental searches to interpret lower lifetime limits in terms of limits on neutrino masses. 
Here we include results from the quasi-particle random-phase approximation (QRPA) \cite{Simkovic2013,Mustonen2013,Hyvarinen2015, Fang2018, Simkovic2018, Jokieniemi2021}, the nuclear shell-model (NSM) \cite{Menendez2016, Horoi2016, Coraggio2020, Jokieniemi2021}, the interacting-boson model (IBM) \cite{Barea2015,Deppisch2020}, both relativistic and non-relativistic energy density functional theory (EDF)~\cite{Rodriguez2010,Vaquero2014,Yao2015,Song2017}, and a hybrid approach combining the NSM using the generalized contact formalism (GCF) with variational Monte-Carlo results in light nuclei to fix short-range correlations~\cite{Weiss2022}.
 
Several attempts have been made to estimate the short-range contributions within these models by taking the charge-independence-breaking coupling constant of the nuclear Hamiltonian as the coupling constant for the contact operator. 
Since the sign of this coupling is unknown, there are two possible bands for these NMEs. 
Nevertheless, first results have been obtained with QRPA and NSM~\cite{Jokieniemi2021} as well as the GCF formalism~\cite{Weiss2022}. 
As seen in Fig.~\ref{fig:model_compare}, ab initio results increase on the order of 60-90\% when including the contact term, still lie at the lower end of NME values with a significantly smaller spread from starting NN+3N forces. 
While work remains to more robustly assess EFT truncation uncertainties, our results appear to strongly disfavour the larger NMEs obtained with particular phenomenological models.

\begin{figure}[t]
    \centering
    \includegraphics[width=\linewidth]{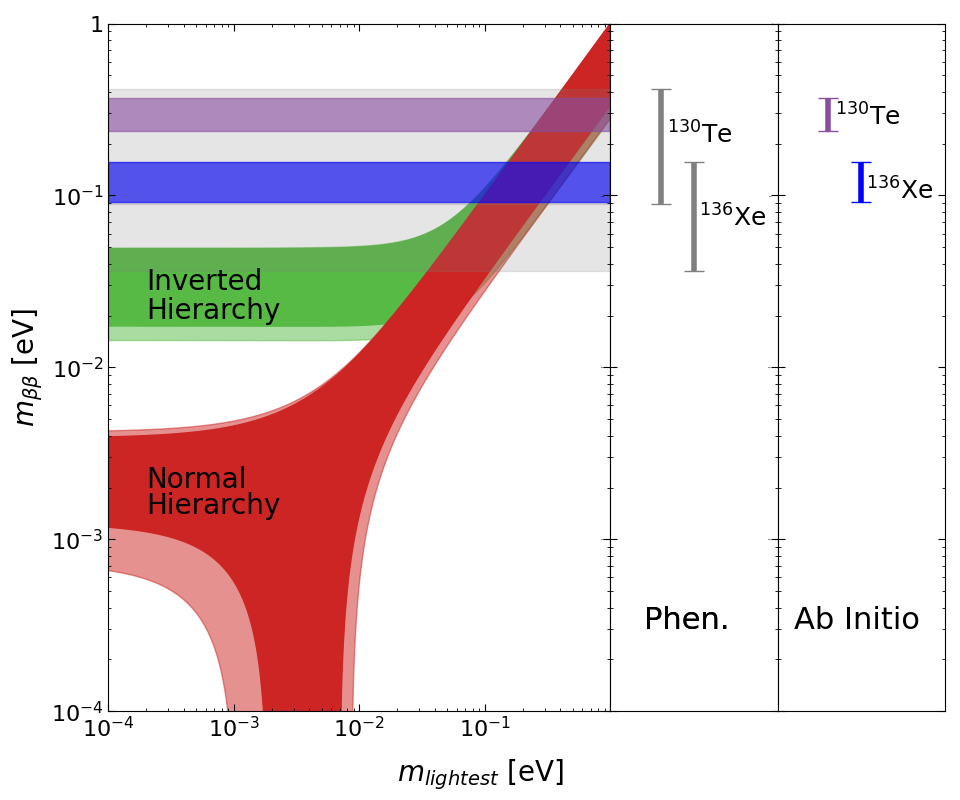}
    \caption{Effective neutrino mass, $m_{\beta \beta}$, extracted from current experimental limits in $^{130}$Te and $^{136}$Xe using phenomenological or ab initio NMEs from Fig.~\ref{fig:model_compare}, compared to the allowed phase-space for both the normal and inverted hierarchies. 
    Lighter shades of the allowed phase-space indicate the $3\sigma$ error on the neutrino oscillation parameters taken from~\cite{Adams2022}. 
    }
    \label{fig:lobster}
\end{figure}

To interpret the implications for neutrino masses, the NME connects a given $0\nu\beta\beta$ decay lifetime limit to the effective neutrino mass through the following relation:
\begin{equation*}
    [T^{0\nu\beta\beta}_{1/2}]^{-1}=G^{0\nu}|M^{0\nu\beta\beta}|^2 \bigg(\frac{\langle m_{\beta\beta}\rangle}{m_e}\bigg) ^2,
\end{equation*}
where $T^{0\nu\beta\beta}_{1/2}$ is the half-life of the decay, $G^{0\nu}$ a well-established phase-space factor~\cite{Stoica2019}, $M^{0\nu\beta\beta}$ the NME, and  $m_{\beta\beta}$ is the effective Majorana mass of the neutrino.  
We relate $m_{\beta\beta}$ to the neutrino mass eigenstates, $m_k$, via $\langle m_{\beta\beta}\rangle =  \sum_k U_{ek}^2m_k$, where $U_{ek}$ are the elements of the Pontecorvo–Maki–Nakagawa–Sakata matrix, connecting neutrino mass and flavour eigenstates. 
While the absolute scale of the mass eigenstates is unknown, we know $m_1$ and $m_2$ have a similar squared masses in addition to the squared mass difference between these two and $m_3$~\cite{DeSalas2018}.
This creates two different scenarios:\ the normal hierarchy (NH), where $m_3$ is the heaviest; and the inverted hierarchy (IH), where $m_3$ is the lightest. 
Using the values of the oscillation parameters \cite{Adams2022}, we can constrain the allowed effective mass of the neutrino, $m_{\beta \beta}$, as a function of the lightest neutrino state, $m_{\mathrm{lightest}}$, for both hierarchies. 

In Fig.~\ref{fig:lobster}, we compare limits on the effective neutrino mass to allowed values for both hierarchies, extracted with either conventional phenomenological NMEs or our ab initio results (using accepted $G^{0\nu}$ values~\cite{Stoica2019}). 
Here we take the half-life limits from CUORE~\cite{Adams2022} ($T^{0\nu\beta\beta}_{1/2} > 2.2 \times 10^{25}$yr) and KamLAND-Zen~\cite{KamLand2023} ($T^{0\nu\beta\beta}_{1/2} > 2.3 \times 10^{26}$yr), the current best experimental limits for $^{130}$Te and $^{136}$Xe, respectively. 
We see that with ab initio NMEs, not only is the uncertainty significantly smaller, but the experimental reach is reduced by nearly an order of magnitude.
Our results suggest that, except for the quasi-degenerate region where neutrino masses are nearly the same for both hierarchies, most of the allowed effective neutrino mass phase space has not yet been probed by any current experiment.
This is in contrast to claims that, with particular phenomenological NMEs, the inverted mass hierarchy has already been partially probed by recent KamLAND-Zen observations~\cite{KamLand2023}. 
Finally, these new results are vital for the strategic planning of next-generation ton-scale searches, which endeavour to completely probe the inverted hierarchy.
With anticipated half-life sensitivities~\cite{Adhikari_2022} on the order of 10$^{28}$yr, given the range of ab initio NMEs presented here, this is unlikely to be achieved with current time and material allocations.

Ab initio nuclear theory provides the most complete account for physics expected to be relevant for NMEs in all $0\nu\beta\beta$ decay nuclei, at once offering a consistent treatment of the new short-range contact contribution, as well as a viable path towards rigorous quantification of theoretical uncertainties. 
We stress, however, that while these results are promising first steps in heavy systems, they do not yet represent final values for the NMEs. 
Further analysis of theoretical uncertainties (similar to recent $^{208}$Pb studies~\cite{Hu2022}) is needed to rigorously assess errors arising from i) the choice of parameters as well as truncations in the expansion of chiral nuclear forces, ii) neglected many-body physics in the IMSRG(2) approximation, and iii) neglected higher-order contribution to the $0\nu\beta\beta$ decay operator. 
With the development of IMSRG-based emulators, this level of EFT uncertainty quantification is already within reach and currently underway. 
Calculations explicitly including higher-order contributions to the matrix elements, while not relying on the closure approximation, could potentially reduce the ab initio uncertainty to the level where discrimination between different proposed $0\nu\beta\beta$ decay mechanisms is possible, in the event of an eventual observation~\cite{Graf2022}.
Nevertheless the values presented here, which lie at the lower end of previous calculations and reduced spread, already have the potential to refine a major obstacle to interpreting current experimental limits on neutrino masses and planning of next-generation searches.

\putbib[XeTe.bib]

\FloatBarrier

\section*{Acknowledgements}

We thank J. Engel, G. Hagen, H. Hergert, B. S. Hu, L. Jokiniemi, J. Men\'endez, P. Navr\'atil, S. Novario, T. Papenbrock, A. Schwenk, N. Shimizu, and J. M. Yao for valuable discussions and K. Hebeler for providing momentum-space inputs for generation of the 3N forces used in this work.
The IMSRG code used in this work makes use of the Armadillo \texttt{C++} library \cite{Sanderson2016, Sanderson2018}.
TRIUMF receives funding via a contribution through the National Research Council of Canada. 
This work was further supported by NSERC under grants SAPIN-2018-00027 and RGPAS-2018-522453, the Arthur B. McDonald Canadian Astroparticle Physics Research Institute, the Canadian Institute for Nuclear Physics, the US Department of Energy (DOE) under contract DE-FG02-97ER41014, the Deutsche Forschungsgemeinschaft (DFG, German Research Foundation) -- Project-ID 279384907 -- SFB 1245, and the European Research Council (ERC) under the European Union’s Horizon 2020 research and innovation programme (Grant Agreement No.\ 101020842).
Computations were performed with an allocation of computing resources on Cedar at WestGrid and Compute Canada.

\end{bibunit}

\begin{bibunit}

\FloatBarrier

\section*{Methods}
\subsection*{Hamiltonian and model space.} Calculations performed in this work are done using the intrinsic Hamiltonian 
\begin{equation}
    H = \sum_{i < j} (T^{ij}+V_{NN}^{ij}) + \sum_{i<j<k} V_{3N}^{ijk}
\end{equation}
where $T$ is the intrinsic kinetic energy, $V_{NN}$ is the nucleon-nucleon (NN) interaction and $V_{3N}$ is the three-nucleon (3N) interaction.

We use interaction from 3 different families of interactions. 
The first family consists of a NN interaction at N3LO~\cite{Entem2003} in chiral order which are evolved at the scale $\lambda$ via  similarity renormalization group (SRG) and an unevolved 3N interaction at N2LO with a non-local regulator with cutoff $\Lambda = 2.0$ fm$^{-1}$ {}~\cite{Hebeler2011}. 
We denote these interaction by EM($\lambda$/$\Lambda$). 
In particular we use interaction at scale $\lambda =1.8$ fm$^{-1}$ namely EM(1.8/2.0). 
3N coupling constants are constrained by the binding energy of $^{3}$H and the charge radius of $^{4}$He. EM(1.8/2.0) in particular is known for being able to reproduce ground state energy for isotope of to $^{100}$Sn, albeit underpredicting nuclear charge radii \cite{Hebeler2011, Simonis2016, Morris2018}. 

We further use the $\Delta$N2LO\textsubscript{GO}(394) Hamiltonian~\cite{Jiang2020} (denoted $\Delta_{GO}$ for simplicity), which accounts for $\Delta$ isobars in its construction. 
This low-cutoff NN+3N interaction uses $A \leq 4$ few-body data and nuclear matter properties to constrain the coupling constants.

Finally, we use the combination of N3LO NN~\cite{Entem2003} and local-non-locally regulated 3N~\cite{Soma2020} at N2LO (denoted by $N3LO_{LNL}$), introduced in Ref.~\cite{Leistenschneider2018}, with consistently SRG-evolved NN and 3N forces. 
The NN forces are taken to be the same as the EM family and the 3N interaction uses a mixture a local and non-local regulator with cutoff of 650 MeV and 500 MeV respectively. 
Coupling constants are constrained with binding energies and half-life of the triton and $^{4}$He.

We truncate the one-body model space for the Hamiltonian with a truncation in $e_{\textrm{max}}$ where $e = 2n+l \leq e_{\textrm{max}}$ where $n$ is the radial quantum number and $l$ is orbital angular momentum. 
We further have to truncate the 3N forces with a cut $e_1 + e_2 + e_3 \leq E_{3\textrm{max}}$ where $E_{3\textrm{max}} = 3\cdot e_{\textrm{max}}$ includes all 3N forces. 

\subsection*{$0\nu\beta\beta$ operators.} 

In this work, we use operators derived using the closure energy to avoid explicit sums over all the possible intermediate states. 
In this approximation, the energy dependence is approximated by an average closure energy $E_c$ to make the operator independent of the intermediate states. 
This way the decay can be seen as a pure two-body operators from the parent nuclei ground state to the daughter nuclei ground state. 
Corrections to this approximation are of order $E_c/q\sim$10\%~\cite{Horoi2013}. 
In an EFT framework, these corrections only appear at sub-leading order~\cite{Cirigliano2018}. We note that results from all other models presented in this work also use this approximation.  

\begin{figure*}[]
\centering
\includegraphics[width=\textwidth]{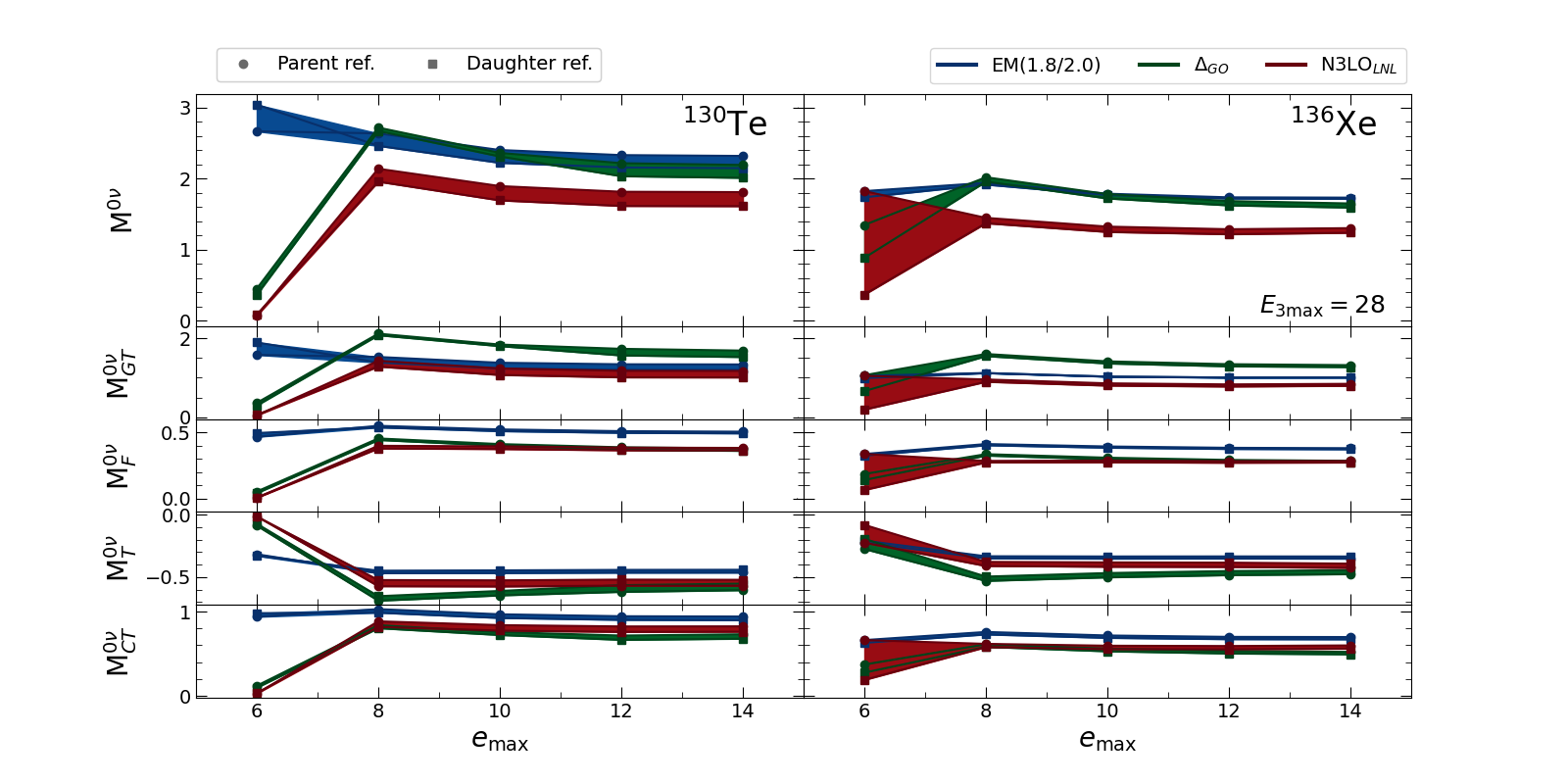}  
\caption{\label{fig:emax}  Convergence in $e_{\textrm{max}}$ for the three interaction used at the maximal value of  $E_{3\textrm{max}} = 28$ used in this work. We further show the reference state dependence of our calculations.}
\end{figure*}

The long range operators can then be written as 
\begin{equation}
    O_{\alpha} (\mathbf{q}) = \frac{R_{Nucl}}{2\pi} \frac{h_\alpha(q)}{q(q+E_c)} S_\alpha(\mathbf{q})\tau_{1}\tau_{2}
\end{equation}
where $\alpha \in [\textrm{GT,F,T}]$, $q$ is the momentum transfer, $R_{Nucl} = 1.2A^{1/3}$~fm is the nuclear radius included to make the NME dimensionless, and $\tau$ is the isospin-raising operator. 
We choose the standard values of $E_c = 12.76$ and $13.06$ MeV for $^{130}$Te and $^{136}$Xe respectively\cite{Tomoda1991}.
The spin-spatial part of the operators $S_\alpha(\mathbf{q})$ is given respectively by
\begin{align}
    &S_F(\mathbf{q}) = \mathds{1}\\
    &S_{GT}(\mathbf{q}) = \mathbf{\sigma_1 \cdot \sigma_2}\\
    &S_{T}(\mathbf{q}) = -\big[3\frac{(\mathbf{\sigma_1\cdot q})(\mathbf{\sigma_2\cdot q})}{q^2} - (\mathbf{\sigma_1 \cdot \sigma_2})\big]
\end{align}
where $\sigma$ is the Pauli matrix acting on spins.

At N2LO, the neutrino potentials $h_\alpha(q)$ are defined using the vector ($g_V$), axial-vector ($g_A$), induced pseudoscalar ($g_P$) and weak-magnetism ($g_M$) coupling constants as 
\begin{align}
h_F(q) &= \frac{g_V^2(q)}{g_V^2(0)},\\
h_{GT}(q) = &\dfrac{g_A^2(q)}{g_A^2(0)}\Bigg[1-\dfrac{2}{3}\dfrac{q^2}{q^2+m_\pi^2}+\dfrac{1}{3}\bigg(\dfrac{q^2}{q^2+m_\pi^2}\bigg)^2\bigg]\nonumber\\
&\hphantom{{}={}}+\dfrac{1}{6}\dfrac{g_M^2(q)}{g_A^2(0)}\dfrac{q^2}{m_N^2},\\ 
h_{T}(q) = &\dfrac{g_A^2(q)}{g_A^2(0)}\Bigg[\dfrac{2}{3}\dfrac{q^2}{q^2+m_\pi^2}-\dfrac{1}{3}\bigg(\dfrac{q^2}{q^2+m_\pi^2}\bigg)^2\bigg]\nonumber\\
&\hphantom{{}={}}+\dfrac{1}{12}\dfrac{g_M^2(q)}{g_A^2(0)}\dfrac{q^2}{m_N^2},
\end{align}
where $m_\pi = 138.039$ MeV is the pion mass, $m_N = 938.919$ MeV is the average nucleon mass and  the coupling constants are regularized to account for nucleon finite size effects as 

\begin{align}
g_V(q) &= g_V(0)\left(1+ q^2/\Lambda^2_V\right)^{-2},\\
g_A(q) &= g_A(0)\left(1+ q^2/\Lambda^2_A\right)^{-2},\\
g_P(q) &= g_A(q^2) \left(\dfrac{2m_p}{ q^2+m^2_\pi}\right),\\
g_M(q) &= g_V(q^2)\left(1+\kappa_1\right).
\end{align}

We use the standard values of $g_V(0)=1$ and $g_A(0)=1.27$ in this work, and isovector anomalous magnetic moment of the nucleon  $\kappa_1 = 3.7$. 
Following Ref.~\cite{Simkovic2009}, we use $\Lambda_V = 850$ MeV for the vector regulator and $\Lambda_A = 1086$ MeV for the axial regulator.

The short range operator corresponds to a simple contact operator which we regularize using a non-local regulator, such that
\begin{equation}
    O_{CT}(p,p') = \frac{R_{Nucl}}{8\pi^3}\bigg(\frac{m_N g_A^2}{4 f_\pi^2}\bigg)^2 e^{-(\frac{p}{\Lambda_{int}})^{2n_{int}}}e^{-(\frac{p}{\Lambda_{int}})^{2n_{int}}}
\end{equation}
where $p$ and $p'$ are the relative momenta of the initial and final states; $f_\pi = 92.2$ MeV is the pion decay constant; and $\Lambda_{int}$ and $n_{int}$ corresponds to the regulator cutoff and regulator power of the interaction used. 
In our case, we have $\Lambda_{int} = 500$ MeV and $n_{int} = 3$ except for the $\Delta_{GO}$ interaction where $\Lambda_{int} = 394$ MeV and $n_{int} = 4$.

\subsection*{VS-IMSRG}

The IMSRG~\cite{Tsukiyama2011, Hergert2016} uses a unitary transformation $U$ to transform the initial Hamiltonian $H$ into a diagonal or block diagonal form, generated as $\tilde{H} = UHU^\dagger$. 
In the specific formulation used in this work, namely the Magnus formulation, the transformation is expressed as the exponential of an anti-hermitian generator $\Omega$ such that $U=e^\Omega$, where $\Omega$ encodes the physics to be decoupled. 
Starting from a single-reference ground state $\ket{\Phi_0}$ obtain via the Hartree-Fock method, we apply a continuous sequence of unitary transformation $U(s)$ to obtain the fully correlated ground state $\ket{\Psi_0}$. 
In this work, we use the IMSRG(2) scheme, meaning that we truncate all operators at the two-body levels, which introduces some error in this otherwise exact method.

In the valence-space (VS) formulation of the IMSRG~\cite{Bogner2014,Stroberg2017,Stro19ARNPS,Miya20lMS}, the generator $\Omega$ is defined to decouple a valence-space Hamiltonian $H_{VS}$ from the rest of the Hilbert space. 
To account for 3N forces inside the valence-space, we utilize ensemble normal ordering at the two-body level (NO2B). We obtain the eigenstates using the KSHELL shell-model code \cite{Shimizu2019}, which performs the valence-space diagonalization. 
We further evolve all operators consistently using the same transformations that were used for the Hamiltonian, namely $\tilde{O} = e^\Omega O  e^{-\Omega}$.

The valence-space used for both nuclei consists of the $0g_{7/2}$,$1d_{3/2}$,$1d_{5/2}$,$2s_{1/2}$,$0h_{11/2}$ proton and neutron orbits outside of a $^{100}$Sn core. Considering effects of the variation of the valence-space on the NME would be interesting and is out of the scope of this work due to limitation with respect to the size of the valence-space considered.

We note that the choice of reference state for ensemble normal ordering is ambiguous in our case as both the parent or daughter nuclei or even the intermediate nuclei are valid choices and choosing a different reference state results in small differences in the final results due to the NO2B truncation. To account for this error, we present results for both the parent and daughter nuclei. These differences are expected to decrease when going to the IMSRG(3) truncation scheme where all operators are truncated at the 3-body level, and will need to be investigated further in future work. 

\subsection*{$E_{3\textrm{max}}$ extrapolation.}
To perform the extrapolation in $E_{3\textrm{max}}$ so that we can consider all 3N forces for a given value of $e_{\textrm{max}}$, we use the extrapolation described in Ref.~\cite{Miyagi2022} for energies. 
Assuming the HF energies are well converged with respect to $E_{3\textrm{max}}$, the extrapolation is given by 
\begin{equation}
    E \approx A \gamma_{\frac{2}{n}}\bigg[\bigg(\frac{E_{3\textrm{max}}-\mu}{\sigma}\bigg)^n\bigg] + C
\end{equation}
where $\gamma_{s}(x)$ is the incomplete gamma function, $n$ is a parameter that depends on the form of the interaction, generally taken to be $n=4$, and $A$, $\mu$, $\sigma$ and $C$ are parameters that need to be fitted. 
This extrapolation is also expected to hold for one-body operators. 
In our case, we use it for two-body operators and verify that it still captures the convergence in $E_{3\textrm{max}}$ correctly. 
To do so we try different value of $n$ for the extrapolation, namely $n=2,4,6$ and find no significant changes to our results, and therefore, we present only the results with $n=4$. 
We further consider the extrapolation using only $E_{3\textrm{max}} \in [16,18,20,22,24]$ for $e_{\textrm{max}}=14$ and compare the extrapolation with our computed value at $E_{3\textrm{max}} = 26,28$ and find them to be within the error of the fit, which we obtain by taking $10^4$ samples from the covariance matrix. 
Therefore, we expect this extrapolation to hold for this case even if we are working with a two-body operator. We note that this might not be the case for other two-body operators and should be verified.

\putbib[methods_refs.bib]
\end{bibunit}

\setcounter{figure}{0}
\setcounter{table}{0}
\renewcommand{\figurename}{Extended Data Figure}
\renewcommand{\tablename}{Extended Data Table}
\renewcommand{\thefigure}{\arabic{figure}}
\renewcommand{\thetable}{\arabic{table}}

\end{document}